\documentclass[11pt,eqs]{article}
\usepackage{latexsym,amsmath,amstext,amssymb}

\def\cleq{\setcounter{equation}{0}}

\title{
Canonical approach to the closed string non-commutativity
\thanks{Work supported in part by
the Serbian Ministry of Education, Science and Technological Development, under contract No. 171031.}}
\author{Lj. Davidovi\'c\thanks{e-mail: ljubica@ipb.ac.rs} ,
B. Nikoli\'{c}\thanks{e-mail: bnikolic@ipb.ac.rs}\,
and B. Sazdovi\'c\thanks{e-mail: sazdovic@ipb.ac.rs}\\
{\it Institute of Physics,}\\
{\it University of Belgrade,}\\
{\it 11001 Belgrade, P.O.Box 57, Serbia}
}

\begin{document}
\maketitle

\begin{abstract}
We consider the closed string
moving in the weakly curved background and its totally T-dualized background.
Using T-duality transformation laws,
we find the structure of the Poisson brackets in the T-dual space corresponding to the fundamental Poisson brackets in the original theory. From this structure we obtain 
that the commutative original theory is equivalent to the
non-commutative T-dual theory, whose Poisson brackets are proportional to the
background fluxes times winding and  momenta numbers.
The non-commutative  theory of the present article is more nongeometrical then T-folds and 
in the case of three space-time dimensions 
corresponds to the nongeometric space-time with $R$-flux.
\end{abstract}
%%%%%%%%%%%%%%%%%%%%%%%%%%%%%%%%%%%%%%%%%%%%%%%%%
\section{Introduction}
\setcounter{equation}{0}

It is well known that the open string endpoints,
attached to $Dp$-brane,
are non-commutative \cite{k1,k2}.
The non-commutativity
is implied by the fact that,
on the solution of boundary conditions
%the solution of the boundary conditions gives
the initial coordinate is given as a linear combination of the effective coordinate and the effective momentum,
which have the nonzero Poisson bracket (PB).
In the constant background case,
the coefficient in front of momenta is proportional to the Kalb-Ramond field $B_{\mu\nu}$, whose presence is crucial in gaining the non-commutativity.

The closed string does not have endpoints and in the flat space the boundary conditions are satisfied automatically.
But,
to understand the closed string non-commutativity, we are going to use
the similar explanation
as in the open string case.
We will express the closed string coordinates in terms of the coordinates and momenta of
some other space.
The relation between different spaces will be established using the
T-duality transformations.

The T-dualization along isometry directions,
and the construction of T-dual theory
was first realized through Buscher procedure
\cite{buscher}.
The procedure is in fact a localization of the translation
invariance symmetry, in which beside the covariantization of derivatives one adds the Lagrangian multiplier term to the action which
insures the physical equivalence of the initial and the T-dual theory.

In the flat space, T-duality relates
$\sigma$-derivatives of the coordinates of the original theory with the momenta of its T-dual theory,
and vice versa.
As the momenta of the original theory are taken to be commutative it follows that the coordinates commute as well.
So, in the flat space there is no non-commutativity
of the closed string T-dual coordinates.
This is in agreement with the fact that T-duality is canonical transformation in the flat space,
and the fact that PB's are invariant under such transformations.
%so as the initial coordinates commute, the T-dual coordinates commute as well.

The closed string non-commutativity was first observed
in the paper \cite{lust}, and investigated further in \cite{ALLP,ALLP2,L},
where it was found that
the commutators of the coordinates are proportional to
% the non-commutativity parameter depends on
the flux and the winding number. 

Let us briefly describe the result of Ref. \cite{ALLP} following
its notation. After the $T_1$-dualization along coordinate
$X^{1}$, one obtains the twisted torus with coordinates $Y^a,
(a=1,2,3)$ and $f$-flux. After additional $T_2$-dualization along
$X^{2}=Y^{2}$ one obtains the nongeometric background with
coordinates $Z^a$ and $Q$-flux. Using the standard Buscher
prescription one can not perform $T_3$-dualization along the
coordinate $X^{3}=Y^{3}=Z^{3}$ because the Kalb-Ramond field
$B_{ab}$ depends on $Z^{3}$. But it is argued in Refs. \cite{nongeo1,ALLP}
that $T_3$-dualization leads to the nongeometric background with
R-flux configuration and coordinates $W^{a}$ presented in the
T-duality chain
\begin{eqnarray}
&&H_{abc},\,X^{a}
\xrightarrow{T_1}
 f^{a}_{\ bc},\,Y^{a}
\xrightarrow{T_{2}}
Q^{ab}_{\ c},\,Z^{a}
\xrightarrow{T_{3}}
R^{abc},\,W^{a}.
\end{eqnarray}
In the paper \cite{ALLP}, the non-commutativity of the nongeometric background ($Z^a$ with $Q$-flux) has been obtained using its
$T_{2}$-duality connection  $Z^{a}=Z^{a}(Y^{a})$ with geometric background (twisted torus with $Y^a$ and $f$-flux).

In our paper \cite{DS}, we performed generalized Buscher's
T-dualization procedure along all the coordinate directions. It corresponds to   $T=T_1\circ T_2\circ \dots \circ  T_{D}$ -duality relation
$y_\mu = y_\mu (x^\mu)$ connecting  the beginning and the end of the T-duality chain
\begin{eqnarray}\label{eq:chain2}
&&H_{\mu \nu \rho},\,x^{\mu}
\xrightarrow{T_1}
(f_1)_{\mu \nu \rho},\,x_1^{\mu}
\xrightarrow{T_{2}}
(f_2)_{\mu \nu \rho},\,x_2^{\mu}
\xrightarrow{T_{3}} \dots  \xrightarrow{T_{D}}
(f_D)_{\mu \nu \rho},\,x_D^{\mu}=y_\mu  ,
\end{eqnarray}
where $(f_{i})_{\mu\nu\rho}$ and $x^\mu_{i},\,(i=1,2,\cdots,D)$ are fluxes and the coordinates of the corresponding configuration. In $D$-dimensional space-time it is possible to perform T-duality along any subset of coordinates.
For simplicity, in the present article we will T-dualize all the directions. The general case will be published separately.

We considered the bosonic string moving in
the background with the constant metric
$G_{\mu\nu}=const$ and the linear Kalb-Ramond field
$B_{\mu\nu}=b_{\mu\nu}+\frac{1}{3}B_{\mu\nu\rho}x^\rho$, where
the field strength of the Kalb-Ramond field
$B_{\mu\nu\rho}$ is infinitesimally small (for more details see the introductory part of Section 2). The obtained T-dual theory is
of the same form as the initial theory, so that the T-dual string moves
in the T-dual background but in the doubled space given by the coordinates $y_\mu$, $\tilde{y}_\mu$.
The dual coordinates satisfy the following conditions
 $\dot{y}_\mu=\tilde{y}^\prime_\mu$, ${y}^\prime_\mu=\dot{\tilde{y}}_\mu$.
The improvement, in comparison to the standard Buscher procedure,
is the covariantization of the coordinates  $x^\mu$.
In fact, because $x^\mu$ is gauge dependent, it is replaced by the
gauge invariant expression
$\Delta x^\mu_{inv}=\int d\xi^\alpha D_{\alpha}x^\mu$.
As pointed out in \cite{nongeo1}, the T-dual background of the present paper,
is of the "new class that is even more nongeometrical than $T$-folds".
Unlike the T-folds, this background is not standard manifold even locally.
In our formulation, this
stems from the fact that the argument of the background fields
$\Delta x^\mu_{inv}$ is the line integral.

In the canonical formalism, the T-dual variables can be expressed in terms of the original ones
in the simple form
$y^\prime_\mu\cong
\frac{1}{\kappa}\pi_\mu-\beta^{0}_\mu[x]$
and
$
^\star\pi^\mu\cong
\kappa x^{\prime\mu}+\kappa^{2}\theta_{0}^{\mu\nu}\beta^{0}_\nu[x].
$
The infinitesimal expression $\beta^{0}_\mu$ is the improvement in comparison to the flat background case.
Because the coordinates and momenta of the original theory do not commute,
$\beta^{0}_\mu$ is the source of the closed string noncommutativity.

% The improvement, in comparison to the standard Buscher procedure,
% is the covariantization of the coordinates  $x^\mu$.
% In fact, because $x^\mu$ is gauge dependent, it is replaced by the
% gauge invariant expression
% $\Delta x^\mu_{inv}=\int d\xi^\alpha D_{\alpha}x^\mu$.
% As pointed out in \cite{DH}, the T-dual background of the present paper, is more nongeometric than $T$-folds.
% Unlike the T-folds, this background is not standard manifold even locally.
% In our formulation, this is obvious because the argument of the background fields
% $\Delta x^\mu_{inv}$ is the line integral.

We will follow the main idea of Ref. \cite{ALLP},
using the T-duality transformation laws between the T-dual backgrounds
in order to study the non-commutativity of the coordinates.
In the paper \cite{ALLP},
the $T_{2}$-duality connects coordinates $Z^{a}=Z^{a}(Y^{a})$ of the nongeometric background ($Z^a$ with $Q$-flux) and the geometric background
(twisted torus with $Y^a$ and $f$-flux). We performed T-dualization procedure along all the coordinates, and obtained the T-duality transformation
$y_\mu = y_\mu (x^\mu)$ of the locally nongeometric background
(the end of the chain (\ref{eq:chain2}) with $y_\mu$ and $f_D$-flux) and the geometric background (torus with $H$-flux in the beginning of the chain (\ref{eq:chain2})).
In both approaches it was assumed that the geometric backgrounds
(described by $Y^{a}$ in \cite{ALLP} and by $X^{a}$ in our paper)
have the standard commutation relations.
The PB between $y_\mu$'s is proportional to the flux
$B_{\mu\nu\rho}$ and the winding number $N^\mu$ of the initial theory.  In addition,
we obtain the complete algebra of the T-dual coordinates and momenta in terms of the fluxes.

For $D=3$, the case of the present article corresponds to T-duality  $T=T_1\circ T_2\circ T_{3}$
which connects the
coordinates $W^{a}=W^{a}(X^{a})$ of the nongeometric background ($W^a$ with $R$-flux) and the geometric background (torus with $X^a$ and $H$-flux).
In comparison to Ref. \cite{ALLP}, this procedure  contains
one $T$-dualization more, $T_3$-dualization along the coordinate $X^{3}=Y^{3}=Z^{3}$,
which can not be done using the standard Buscher prescription because the Kalb-Ramond
field $B_{ab}$ depends on $Z^{3}$.
So, in terms of Ref.\cite{ALLP}, we obtained the non-commutativity of the nongeometric background, with R-flux configuration.
This background does not look like the conventional space even locally.
%which is more nongeometric than T-folds.

% The PB between $y_\mu$'s is proportional to the flux
% $B_{\mu\nu\rho}$ and the winding number $N^\mu$ of the initial theory.  In addition,
% we obtain the complete algebra of T-dual coordinates and momenta in terms of the fluxes.

At the end we give three appendices. In the first one we derive in detail the expression for the dual momentum
${}^\star \pi^\mu$, while in the second one we make a list of fluxes used in the paper. The third appendix contains
the mathematical details about transition from PB $\{\Delta X,\Delta Y\}$ to PB $\{X,Y\}$.

%%%%%%%%%%%%%%%%%%%%%%%%%%%%%%%%%%%%%%%%%%%%%%%%%%%%%%%%%%%%%%%%%%%%%%%%%%%%%%%%%%%%%%%%%%%%%%%%%%%%%
\section{Bosonic string in the weakly curved background and its T-dual picture}
\cleq

Let us consider the closed string moving
in the $D$-dimensional space -time,
in the coordinate $x^\mu(\tau,\sigma),\,\mu=0,\cdots,D-1$ dependent background,
described by the action
\begin{equation}\label{eq:action1}
S[x] = \kappa \int_{\Sigma} d^2\xi\
\partial_{+}x^{\mu}
\Pi_{+\mu\nu}[x]
\partial_{-}x^{\nu}.
\end{equation}
We suppose that all the coordinates are compact with the radii $R_\mu$.
The background is defined by the space-time metric $G_{\mu\nu}$ and the antisymmetric Kalb-Ramond field $B_{\mu\nu}$
\begin{equation}\label{eq:pipm}
\Pi_{\pm\mu\nu}[x]=
B_{\mu\nu}[x]\pm\frac{1}{2}G_{\mu\nu}[x].
\end{equation}
The light-cone coordinates are
\begin{equation}
\xi^{\pm}=\frac{1}{2}(\tau\pm\sigma),
\qquad
\partial_{\pm}=
\partial_{\tau}\pm\partial_{\sigma},
\end{equation}
and the action is given in the conformal gauge
(the world-sheet metric is taken to be $g_{\alpha\beta}=e^{2F}\eta_{\alpha\beta}$).

The world-sheet conformal invariance is required,
as a condition of having a consistent theory on the the quantum level \cite{P,S}.
This results in the following space-time equations for the background fields
\begin{equation}\label{eq:stem}
R_{\mu \nu} - \frac{1}{4} B_{\mu \rho \sigma}
B_{\nu}^{\ \rho \sigma}=0,\quad
D_\rho B^{\rho}_{\ \mu \nu} = 0,
\end{equation}
in the lowest order in slope parameter $\alpha^\prime$ and for the constant dilaton field $\Phi=const$.
Here
\begin{equation}\label{eq:fskr}
B_{\mu\nu\rho}=\partial_\mu B_{\nu\rho}
+\partial_\nu B_{\rho\mu}+\partial_\rho B_{\mu\nu}
\end{equation}
is the field strength of the field $B_{\mu \nu}$, and
$R_{\mu \nu}$ and $D_\mu$ are Ricci tensor and
the covariant derivative with respect to the space-time metric.

We will consider
the weakly curved background \cite{ALLP,DS,wcb,DS1},
defined by
\begin{eqnarray}\label{eq:wcb}
G_{\mu\nu}[x]&=&const,
\nonumber\\
B_{\mu\nu}[x]&=&b_{\mu\nu}+h_{\mu\nu}[x]=b_{\mu\nu}+
\frac{1}{3}B_{\mu\nu\rho}x^\rho,\qquad b_{\mu\nu},B_{\mu\nu\rho}=const.
\end{eqnarray}
Here, the constant
$B_{\mu\nu\rho}$ is infinitesimally small which, according to \cite{lust,ALLP,L}, means that we will assume
that $D$ dimensional torus is so large that for any $\mu,\nu,\rho$ 
\begin{equation}\label{eq:binf}
\frac{B_{\mu\nu\rho}}{R_\mu R_\nu R_\rho} \ll 1\, ,
\end{equation}
where $R_\mu (\mu=0,1,\dots D-1)$ are the radii of the
torus. For simplicity we will take $R_0=R_1=\dots =R_{D-1}$ and rescale background fields according to App.A of Ref. \cite{ALLP}.
%This additional term in the Kalb-Ramond field will cause
%the closed string non-commutativity.
The background (\ref{eq:wcb}) is the solution of (\ref{eq:stem}) in the
first order in $B_{\mu\nu\rho}$ approximation of the closed string theory (\ref{eq:action1}).

%%%%%%%%%%%%%%%%%%%%%%%%%%%%%%%%%%%%%%%%%%
\subsection{T-dual bosonic string}

The T-dualization of the
closed string theory in the weakly curved background was a subject of investigation in \cite{DS}.
There we presented  the
T-dualization procedure performed along all the coordinates,
in a background which depends on these coordinates.
Here we will give a short overview of the most important results.

The T-dual picture of the theory
is given by
\begin{equation}\label{eq:dualna}
^{\star}S[y]=
\kappa
\int d^{2}\xi\
\partial_{+}y_\mu
\,^\star \Pi_{+}^{\mu\nu}\big[\Delta  V[y]\big]\,
\partial_{-}y_\nu
=\,
\frac{\kappa^{2}}{2}
\int d^{2}\xi\
\partial_{+}y_\mu
\Theta_{-}^{\mu\nu}\big[\Delta  V[y]\big]
\partial_{-}y_\nu,
\end{equation}
with
\begin{eqnarray}\label{eq:dualback}
{\Theta}^{\mu\nu}_{\pm}\equiv -\frac{2}{\kappa}
(G^{-1}_{E}\Pi_{\pm}G^{-1})^{\mu\nu}=
{\theta}^{\mu\nu}\mp \frac{1}{\kappa}(G_{E}^{-1})^{\mu\nu},
\quad G_{E\mu\nu}\equiv G_{\mu\nu}-4(BG^{-1}B)_{\mu\nu}.
\end{eqnarray}
The dual background fields defined in analogy with (\ref{eq:pipm}) as ${}^\star \Pi_{\pm}^{\mu\nu}={}^\star B^{\mu\nu}\pm\frac{1}{2}{}^\star G^{\mu\nu}$, have the form
\begin{equation}
^\star G^{\mu\nu}\big[\Delta  V[y]\big]=
(G_{E}^{-1})^{\mu\nu}\big[\Delta  V[y]\big],
\quad
^\star B^{\mu\nu}\big[\Delta  V[y]\big]=
\frac{\kappa}{2}
{\theta}^{\mu\nu}\big[\Delta  V[y]\big]\, .
\end{equation}
Using the terminology introduced in the open string case, they are equal to  the inverse of the effective metric $G^{E}_{\mu\nu}$ and
proportional to the non-commuta\-ti\-vi\-ty
parameter $\theta^{\mu\nu}$.
Their argument is given by
\begin{equation}\label{eq:deltav}
\Delta V^\mu[y]
=-\kappa\theta_{0}^{\mu\nu}\Delta y_\nu
+(g^{-1})^{\mu\nu}\Delta{\tilde{y}}_\nu,
\end{equation}
where
\begin{equation}\label{eq:Deltaxy}
\Delta y_\mu=\int_{P}(d\tau \dot{y}_\mu+d\sigma y^\prime_\mu)=y_\mu(\xi)-y_\mu(\xi_{0}),
\quad
\Delta{\tilde{y}}_\mu=\int_{P}(d\tau y^\prime_\mu+d\sigma \dot{y}_\mu),
\end{equation}
and
\begin{equation}
g_{\mu\nu}=G_{\mu\nu}-4(bG^{-1}b)_{\mu\nu},\quad
{\theta}^{\mu\nu}_{0}
=-\frac{2}{\kappa}
(g^{-1}bG^{-1})^{\mu\nu},
\end{equation}
are constant finite parts of the effective metric and the non-commutativity parameter.
The variable $\Delta\tilde{y}_\mu$ is path independent on the zeroth order equation of motion.
The T-dual theory is defined in the doubled space, defined by two coordinates
$y_\mu$ and $\tilde{y}_\mu$,
related by
expressions
$\dot{y}_\mu=\tilde{y}^\prime_\mu$, ${y}^\prime_\mu=\dot{\tilde{y}}_\mu$.

%%%%%%%%%%%%%%%%%%%%%%%%%%%%%%%%%%%%%%%%%%%%%
\subsection{Transformation laws}

The T-duality transformation
connecting the variables of the closed string theory in the weakly curved background
and its T-dualized  string theory is  \cite{DS}
\begin{equation}\label{eq:xcong}
\partial_{\pm}x^\mu\cong
-\kappa\Theta^{\mu\nu}_{\pm}[\Delta V]
\Big{[}
\partial_{\pm} y_\nu
\pm 2\beta^{\mp}_{\nu} [V]
\Big{]},
\end{equation}
with
\begin{equation}\label{eq:beta}
\beta^\pm_\mu[x]=\frac{1}{2}(\beta^0_\mu\pm\beta^1_\mu)=
\mp\frac{1}{2}
h_{\mu\nu}[x]\partial_{\mp} x^\nu,\,
\beta^0_\mu[x]=h_{\mu\nu}[x]x'^\nu,\,
\beta^1_\mu[x]=-h_{\mu\nu}[x]\dot x^\nu\, .
\end{equation}

From (\ref{eq:xcong}) we can find the transformation law for
$\dot{x}^\mu$ and $x^{\prime\mu}$
\begin{subequations}\label{eq:relx1}
\begin{eqnarray}\label{eq:dotx}
\dot{x}^\mu&\cong&
-\kappa \theta^{\mu\nu}[\Delta V]\dot y_\nu+(G_E^{-1})^{\mu\nu}[\Delta V]y^\prime_\nu+(g^{-1})^{\mu\nu}\beta^0_\nu[V]
+\kappa \theta_0^{\mu\nu}\beta^1_{\nu}[V]
\\\label{eq:xprime1}
x^{\prime\mu}&\cong&
(G_E^{-1})^{\mu\nu}[\Delta V]\dot y_\nu
-\kappa\theta^{\mu\nu}[\Delta V]y^\prime_\nu
-\kappa\theta_0^{\mu\nu}\beta^0_\nu[V]
-(g^{-1})^{\mu\nu}\beta^1_\nu[V]\, .
\end{eqnarray}
\end{subequations}

Using the expression for the
canonical momentum of the original theory
\begin{equation}\label{eq:mp}
\pi_\mu=\frac{\delta S}{\delta \dot{x}^\mu}=
\kappa\Big[G_{\mu\nu}{\dot{x}}^\nu-2B_{\mu\nu}[x]x^{\prime\nu}\Big],
\end{equation}
and T-dual canonical momentum
\begin{equation}\label{eq:dualmom}
^\star\pi^\mu
=\frac{\delta\, ^\star S}{\delta \dot{y}_\mu}=
\kappa(G^{-1}_{E})^{\mu\nu}\big[\Delta  V[y]\big]
\dot{y}_\nu
-\kappa^{2}\theta^{\mu\nu}\big[\Delta  V[y]\big]
y^\prime_\nu
-\kappa  (g^{-1})^{\mu\nu}\beta^1_\nu\big[V[y]\big],
\end{equation}
derived in App. \ref{sec:dm},
we can rewrite the above transformations in the canonical form
\begin{subequations}\label{eq:relx}
\begin{eqnarray}\label{eq:dotx1}
\label{eq:xprime}
x^{\prime\mu}&\cong&
\frac{1}{\kappa}\,^\star\pi^\mu
-\kappa\theta_{0}^{\mu\nu}\beta^{0}_\nu[V],
\\
\pi_\mu&\cong&
\kappa y^\prime_\mu+\kappa\beta^{0}_\mu[V],
\end{eqnarray}
\end{subequations}
with $\beta^{0}_\mu[V]$ defined in (\ref{eq:beta}).
It is shown in Ref. \cite{DS} that the T-dual of the T-dual action is the original one.
The corresponding T-dual transformation
of the variables law
is the inverse of
(\ref{eq:xcong})
\begin{equation}\label{eq:ct}
\partial_{\pm}y_\mu\cong
-2\Pi_{\mp\mu\nu}[\Delta x]\partial_{\pm}x^\nu\mp
2\beta^{\mp}_{\mu}[x],
\end{equation}
and so  the transformation laws for
$\dot{y}_\mu$ and $y^\prime_\mu$
are equal to
\begin{subequations}\label{eq:rely}
\begin{eqnarray}\label{eq:doty}
\dot{y}_\mu&\cong&-2B_{\mu\nu}[x]\dot x^\nu
+G_{\mu\nu}x^{\prime\nu}+\beta^1_\mu[x]\, ,\\
\label{eq:yprime}
y'_\mu&\cong&
G_{\mu\nu}\dot x^\nu-2B_{\mu\nu}[x]x^{\prime\nu}-\beta^0_\mu[x]\,.
\end{eqnarray}
\end{subequations}

Using (\ref{eq:mp}) and (\ref{eq:dualmom})
we obtain the canonical form of the T-dual transformations
\begin{subequations}\label{eq:relycan}
\begin{eqnarray}
\label{eq:yprimecan}
y^\prime_\mu&\cong&
\frac{1}{\kappa}\pi_\mu-\beta^{0}_\mu[x],
\\
\label{eq:starp}
^\star\pi^\mu&\cong&
\kappa x^{\prime\mu}+\kappa^{2}\theta_{0}^{\mu\nu}\beta^{0}_\nu[x].
\end{eqnarray}
\end{subequations}
In the zeroth order one has $x^{(0)\mu}\cong V^{\mu}$,
and it is easy to see that
(\ref{eq:relycan}) is inverse of (\ref{eq:relx}).

Because the T-dual theory is defined in the doubled space,
we will need the canonical expression for $\tilde{y}^\prime_\mu=\dot{y}_\mu$.
Using (\ref{eq:doty}) and (\ref{eq:mp}), we obtain
\begin{equation}\label{eq:ytprime}
\tilde{y}^\prime_\mu\cong
-\frac{2}{\kappa}\Big{(}B[\Delta x]
+\frac{1}{2}h[x]\Big{)}_{\mu\nu}(G^{-1})^{\nu\rho}\pi_\rho
+\Big{(}G^{E}[\Delta x]
-2h[x]G^{-1}b
\Big{)}_{\mu\nu}x^{\prime\nu}.
\end{equation}

%%%%%%%%%%%%%%%%%%%%%%%%%%%%%%%%%%%%%%%%%

\section{Non-commutativity relations
between canonical variables}\label{sec:nc}
\cleq

We want to establish the relation between the Poisson structures
of the original and T-dual theory.
The initial theory is the geometric one,
described by the canonical variables $x^\mu$ and $\pi_\mu$.
So, we choose the standard form of the PB's in the original space,
which are
\begin{equation}
\{x^\mu(\sigma),\pi_\nu(\bar{\sigma})\}=\delta_\nu^\mu\delta(\sigma-\bar{\sigma}),\quad
\{x^\mu(\sigma),x^\nu(\bar\sigma)\}=0,\quad
\{\pi_\mu(\sigma),\pi_\nu(\bar{\sigma})\}=0.
\end{equation}
The T-dual theory is the nongeometric one,
defined in the doubled space, with two coordinates $y_\mu$ and $\tilde{y}_\mu$,
connected by relations
 $\dot{y}_\mu=\tilde{y}^\prime_\mu$, ${y}^\prime_\mu=\dot{\tilde{y}}_\mu$.
Using the T-duality transformation laws,
we search for the corresponding Poisson structure in T-dual theory i.e. the expressions for the
PB's between the T-dual string coordinates
$y_\mu(\sigma)$, $\tilde{y}_\mu(\sigma)$ and momenta $^\star\pi^\mu(\sigma)$.
This is done considering
the brackets between
\begin{eqnarray}\label{eq:DY}
\Delta Y_\mu(\sigma,\sigma_{0})&=&
\int_{\sigma_{0}}^\sigma d\eta\,Y^{\prime}_{\mu}(\eta)=
Y_\mu(\sigma)-Y_\mu(\sigma_{0}),
\end{eqnarray}
$Y_\mu=y_\mu,\tilde{y}_\mu$
and calculating the equal time commutators.
The fact that T-dual coordinates
under T-duality transform to both coordinate and momenta dependent expressions,
enables noncommutativity.
The relation of the form
\begin{equation}\label{eq:formula}
\{X^{\prime}_{\mu}(\sigma),Y^{\prime}_{\nu}(\bar\sigma)\}\cong
K^{\prime}_{\mu\nu}(\sigma)\delta(\sigma-\bar\sigma)
+L_{\mu\nu}(\sigma)\delta^\prime(\sigma-\bar\sigma),
\end{equation}
implies the following relation (derived in the App. \ref{sec:purecoor}) between coordinates
\begin{equation}\label{eq:pbyy}
\{X_\mu(\tau,\sigma),Y_\nu(\tau,\bar\sigma)\}\cong-\left[K_{\mu\nu}(\sigma)-K_{\mu\nu}(\bar\sigma)+L_{\mu\nu}(\bar\sigma)\right]\theta(\sigma-\bar\sigma)\, ,
\end{equation}
where $\theta(\sigma)$ is the step function defined in (\ref{eq:fdelt}).

In the flat space the coordinate dependent part of the Kalb-Ramond field is absent
$h_{\mu\nu}=0$, and consequently $\beta^{0}_\mu=0$. So, from (\ref{eq:yprimecan}) and (\ref{eq:starp}) follows
$y^\prime_\mu\cong\frac{1}{\kappa}\pi_\mu$ and
$^\star \pi^\mu\cong\kappa x^{\prime\mu}$.
Therefore, the PB of the
canonical variables of the T-dual theory
remain the standard ones, the same as in the original theory.
So, the nontrivial infinitesimal expression $\beta^{0}_\mu$,
which exists only in the coordinate dependent backgrounds,
is the source of the
closed string non-commutativity.

Using the transformation laws (\ref{eq:yprimecan}) and (\ref{eq:ytprime}),
we can calculate PB's
$\{y^\prime_\mu,y^\prime_\nu\}$,
$
\{y^\prime_\mu(\sigma),
{\tilde{y}}^\prime_\nu(\bar\sigma)\}$ and
$\{\tilde{y}^\prime_\mu(\sigma),
{\tilde{y}}^\prime_\nu(\bar\sigma)\}$
and express them
in the form of (\ref{eq:formula}) with K and L equal
\begin{enumerate}
\item{$\{y^\prime_\mu,y^\prime_\nu\}$
\begin{equation}\label{eq:yyKL}
K_{\mu\nu}[x]=\frac{3}{\kappa}h_{\mu\nu}[x]=
\frac{1}{\kappa}B_{\mu\nu\rho}x^\rho,
\quad
L_{\mu\nu}=0,
\end{equation}
}

\item{$\{y^\prime_\mu,
{\tilde{y}}^\prime_\nu\}$
\begin{eqnarray}\label{eq:klyyt}
&&K_{\mu\nu}[x,\tilde{x}]=
\frac{3}{\kappa}h_{\mu\nu}[\tilde{x}]
-\frac{6}{\kappa}\Big{[}
h[x]G^{-1}b+bG^{-1}h[x]
\Big{]}_{\mu\nu},
\nonumber\\
&&L_{\mu\nu}[x]=
\frac{1}{\kappa}
g_{\mu\nu}
-\frac{6}{\kappa}\Big{[}
h[x]G^{-1}b+bG^{-1}h[x]
\Big{]}_{\mu\nu},
\end{eqnarray}
with
\begin{equation}\label{eq:tilxprim}
\tilde{x}^{\prime\mu}=\frac{1}{\kappa}(G^{-1})^{\mu\nu}\pi_\nu+2(G^{-1}B)^\mu_{\ \nu}x^{\prime\nu}\, .
\end{equation}
Using (\ref{eq:wcb}) and (\ref{eq:Geff}) expressions (\ref{eq:klyyt}) can be rewritten
in terms of the fluxes
\begin{eqnarray}\label{yytKL}
&&K_{\mu\nu}[x,\tilde{x}]=
\frac{1}{\kappa}B_{\mu\nu\rho}\tilde{x}^\rho
-\frac{3}{2\kappa}\Gamma^{E}_{\rho,\mu\nu}x^\rho,
\nonumber\\
&&L_{\mu\nu}[x]=\frac{1}{\kappa}g_{\mu\nu}
-\frac{3}{2\kappa}\Gamma^{E}_{\rho,\mu\nu}x^\rho,
\end{eqnarray}
}
\item{$\{\tilde{y}^\prime_\mu,
{\tilde{y}}^\prime_\nu\}$
\begin{equation}\label{eq:klqtqt}
K_{\mu\nu}[x]=
\frac{3}{\kappa}h_{\mu\nu}[x]
+\frac{24}{\kappa}\Big{[}bh[x]b\Big{]}_{\mu\nu}
+\frac{6}{\kappa}\Big{[}
h[\tilde{x}]b-bh[\tilde{x}]
\Big{]}_{\mu\nu},
%\frac{1}{\kappa}h_{\mu\nu}[x]-
%\Big{[}g\Big{(}2f_{1}[x]+3f_{2}[\tilde{x}]\Big{)}g
%\Big{]}_{\mu\nu},
\quad
L_{\mu\nu}=0\, .
\end{equation}
In terms of fluxes it becomes
\begin{equation}\label{eq:ytytKL}
K_{\mu\nu}=
-\frac{1}{\kappa}
\Big{[}
B_{\mu\nu\rho}
-6g_{\mu\alpha}Q^{\alpha\beta}_{\ \ \rho}g_{\beta\nu}
\Big{]}x^\rho
+\Big{[}
-\frac{3}{2\kappa}\Big{(}
\Gamma^{E}_{\mu,\nu\rho}-\Gamma^{E}_{\nu,\mu\rho}
\Big{)}
+\frac{4}{\kappa}
B_{\mu\nu\sigma}
(G^{-1}b)^\sigma_{\ \rho}
\Big{]}\tilde{x}^\rho,
\end{equation}
where $\Gamma^{E}_{\nu,\mu\rho}$ and $Q_{\mu\nu\rho}$ are defined in (\ref{eq:Gammaeff}) and (\ref{eq:qu}).

}

\end{enumerate}
For the above values of K and L,
the relation (\ref{eq:pbyy}) gives
\begin{eqnarray}\label{eq:yy}
\{y_\mu(\sigma),y_\nu(\bar\sigma)\}
&\cong&
-\frac{1}{\kappa}B_{\mu\nu\rho}\big{[}x^\rho(\sigma)-x^\rho(\bar\sigma)\big{]}\theta(\sigma-\bar\sigma),
\\\label{eq:yty}
\{y_\mu(\sigma),
{\tilde{y}}_\nu(\bar\sigma)\}&\cong&
-\Big{\{}\frac{1}{\kappa}B_{\mu\nu\rho}\big[\tilde{x}^\rho(\sigma)
-\tilde{x}^\rho(\bar\sigma)\big]
-\frac{3}{2\kappa}\Gamma^{E}_{\rho,\mu\nu}\big[x^\rho(\sigma)-x^\rho(\bar\sigma)\big]
\nonumber\\
&&
+\frac{1}{\kappa}g_{\mu\nu}
-\frac{3}{2\kappa}\,\Gamma^{E}_{\rho,\mu\nu}\,x^\rho(\bar\sigma)
\Big{\}}
\theta(\sigma-\bar\sigma),
\\\label{eq:tyty}
\{\tilde{y}_\mu(\sigma),
{\tilde{y}}_\nu(\bar\sigma)\}&\cong&
-\Big{\{}
-\frac{1}{\kappa}
\Big{[}
B_{\mu\nu\rho}
-6g_{\mu\alpha}Q^{\alpha\beta}_{\ \ \rho}g_{\beta\nu}
\Big{]}
\big[x^\rho(\sigma)-x^\rho(\bar\sigma)\big]
\\
&&
+\Big{[}
-\frac{3}{2\kappa}\Big{(}
\Gamma^{E}_{\mu,\nu\rho}-\Gamma^{E}_{\nu,\mu\rho}
\Big{)}
+\frac{4}{\kappa}
B_{\mu\nu\sigma}
(G^{-1}b)^\sigma_{\ \rho}
\Big{]}
\big[\tilde{x}^\rho(\sigma)-\tilde{x}^\rho(\bar\sigma)\big]
\Big{\}}\theta(\sigma-\bar\sigma)\nonumber\, .
\end{eqnarray}

After two-dimensional reparametrization, the $\sigma$ dependent part takes the form
$$\big[X^\mu(f(\sigma))-X^\mu(f(\bar\sigma))\big]\theta[f(\sigma)-f(\bar\sigma)],$$
where $f(\sigma)$ is monotonically increasing function with properties
$f(0)=0$ and $f(2\pi)=2\pi$.
Therefore,
the PB between different points is not reparametrization invariant.
For fixed points, it can be fit to be arbitrary small,
by the appropriate choice of function $f(\sigma)$.
So, only PB's at the same point are physically significant.

Taking $\sigma=\bar\sigma$ we obtain that all PB's vanish, and consequently, coordinates commute. But, taking $\sigma=\bar\sigma+2\pi$,
in the non-commutativity relation between the dual coordinates $y$'s
(\ref{eq:yy}), we obtain the
{\it closed string non-commutativity relation}
\begin{equation}\label{eq:nc1}
\{y_\mu(\sigma+2\pi),y_\nu(\sigma)\}
\cong
-\frac{2\pi}{\kappa}B_{\mu\nu\rho}N^\rho.
\end{equation}
Here,
$N^\mu=\frac{1}{2\pi}\left[x^\mu(\sigma+2\pi)-x^\mu(\sigma)\right]$
is winding number of the original coordinates. In sec. \ref{sec:lust},
we will compare this relation with the result of
%This result is in agreement with
Ref.\cite{ALLP,L}.

Similarly,
from
(\ref{eq:yty})
and
(\ref{eq:tyty}),
we obtain
\begin{equation}\label{eq:nc2}
\{y_\mu(\sigma+2\pi),\tilde{y}_\nu(\sigma)\}+\{y_\mu(\sigma),\tilde{y}_\nu(\sigma+2\pi)\}
\cong-\frac{4\pi}{\kappa^2}B_{\mu\nu\rho}p^\rho
+\frac{\pi}{\kappa}
\left(3\Gamma^{E}_{\rho,\mu\nu}-8B_{\mu\nu\lambda}b^\lambda{}_\rho\right)N^\rho,
\end{equation}
and
\begin{eqnarray}\label{eq:nc3}
\{\tilde{y}_\mu(\sigma+2\pi),\tilde{y}_\nu(\sigma)\}&\cong&
\frac{2\pi}{\kappa} \left[ -B_{\mu\nu\rho}
-6g_{\mu\alpha}Q^{\alpha\beta}_{\ \
\rho}g_{\beta\nu}+2B_{\mu\nu}{}^\lambda
g_{\lambda\rho}+3\left(\Gamma^E_{\mu,\nu\lambda}-\Gamma^E_{\nu,\mu\lambda}\right)b^\lambda{}_\rho\right]N^\rho\nonumber\\&+&\frac{\pi}{\kappa^2}\left[3\left(\Gamma^E_{\mu,\nu\rho}-\Gamma^E_{\nu,\mu\rho}\right)p^\rho-8B_{\mu\nu\lambda}b^\lambda{}_\rho\right]p^\rho
\, .
\end{eqnarray}
Using
(\ref{eq:tilxprim}) and integrating from $\sigma$ to $\sigma+2\pi$
we have
\begin{equation}\label{eq:wrapping}
\frac{1}{2\pi}\left[\tilde x^\mu(\sigma+2\pi)-\tilde
x^\mu(\sigma)\right]=\frac{1}{\kappa}(G^{-1})^{\mu\nu}p_\nu+2(G^{-1})^{\mu\rho}b_{\rho\lambda}N^\lambda\,
,
\end{equation}
where
\begin{equation}
p_\mu=\frac{1}{2\pi}\int_\sigma^{\sigma+2\pi} d\eta \pi_\mu (\eta)\, .
\end{equation}

To complete the algebra, using the expressions (\ref{eq:relycan})
and (\ref{eq:ytprime}) and after one $\sigma$ integration, we find
that the algebra of $y_\mu$, $\tilde y_\mu$ and ${}^\star \pi^\mu$
is of the following form
\begin{eqnarray}
&&\label{eq:ypi}\left\lbrace y_\mu(\sigma),{}^\star\pi^\nu(\bar\sigma)\right\rbrace\cong
\delta_\mu{}^\nu\delta(\sigma-\bar\sigma)
+\kappa h_{\mu\rho}[x(\sigma)]\theta_0^{\rho\nu}\delta(\sigma-\bar\sigma)
+\kappa h_{\mu\rho}[x^\prime(\bar\sigma)]\theta_0^{\rho\nu}\theta(\sigma-\bar\sigma)\, ,\nonumber
\\ \\
&&\label{eq:typi}\left\lbrace \tilde y_\mu(\sigma),{}^\star\pi^\nu(\bar\sigma)\right\rbrace \cong
\Big{[}
-2bG^{-1}
-3h[x(\sigma)]G^{-1}
-2\kappa bh[x(\sigma)]\theta_{0}
\Big{]}_\mu^{\ \nu}
\delta(\sigma-\bar\sigma)
\nonumber\\
&&\qquad\qquad\qquad\quad -\Big{[}
3h[x^\prime(\bar{\sigma})]G^{-1}+2\kappa b
h[x^\prime(\bar{\sigma})]\theta_{0} \Big{]}_\mu^{\
\nu}\theta(\sigma-\bar\sigma),
\\\nonumber\\
&&\label{eq:pp}
\{^\star\pi^\mu(\sigma),\,^\star\pi^\nu(\bar{\sigma})\}\cong0.
\end{eqnarray}
Note that at the zeroth order one has
$\{y_\mu(\sigma),{}^\star\pi^\nu(\bar\sigma)\}=\delta_\mu^\nu\delta(\sigma-\bar\sigma)$
and $\{\tilde{y}_\mu(\sigma),{}^\star\pi^\nu(\bar\sigma)\}=
-2b_\mu^{\ \nu}\delta(\sigma-\bar\sigma)$,
so both doubled space variables $y_\mu$ and $\tilde{y}_\mu$ have
nontrivial PB with $^\star\pi^\mu$.
%%%%%%%%%%%%%%%%%%%%%%%%%%%%%%%%%%%%%%%%%%%%%%%%%

\section{Comparison with the previous results}\label{sec:lust}
\cleq

Let us mention that the case considered in the present paper is different from that of Ref. \cite{ALLP}.
%{lust}.
In Ref. \cite{ALLP},
%{lust},
the non-commutativity relations in the nongeometric background with $Q$-flux
where established,
which are given in terms of winding numbers on the twisted torus
$N^{3}=\frac{1}{2\pi}\Big{(}
Y^{3}(\sigma+2\pi)-Y^{3}(\sigma)
\Big{)}$.
In the present article,
the non-commutativity
of the nongeometric background, which is not standard even locally and for $D=3$ turns to R-flux background,  was obtained
in terms of the winding numbers on the torus with $H$-flux
$N^\mu=\frac{1}{2\pi}\Big{(}
X^\mu(\sigma+2\pi)-X^\mu(\sigma)
\Big{)}$.

%%%%%%%%%%%%%%%%%%%%%%%%%%%%%%%%%%%%%%%%%%%%%%%%%%%%%%%%%%%%%%%%%%%%%%%%%%%%%%%%%%%%%%%%%%%%%%%%%%%%%%%%%%%
\subsection{The brief overview of the results of Ref.\cite{ALLP}}
Before comparing the results of our paper with those of Ref. \cite{ALLP} let us shortly reexpress result of Ref.\cite{ALLP} using its notation.
From the last identification in Eqs.(2.17) and the first relation in (2.25)
of Ref.\cite{ALLP} it follows that
\begin{equation}
Y^1_H=Y_{0}^2Y_{0}^3+\dots\, .
\end{equation}
Using expression for $G_{ab}(Y_3)$ for twisted torus (Table 1) of Ref. \cite{ALLP} we can find
\begin{equation}
\pi_1=\dot Y^1-HY^3_0\dot Y^2_0\, ,\quad \pi_2=\dot Y^2-HY^3_0 \dot Y^1_0\, ,
\end{equation}
and consequently
\begin{equation}
\pi_{01}=\dot Y_0^1\, ,\quad \pi_{H2}=\dot Y_H^2-Y_0^3\dot Y^1_0=\dot Y_H^2-Y^3_0\pi_{01}\, .
\end{equation}

The $T_2$-duality
along $Y^{2}$,
from the twisted torus to the nongeometric background produces
\begin{equation}
Z^1\cong Y^1=Y^1_0+HY_{0}^2 Y_{0}^3\, ,\quad {Z^2}'\cong \dot Y^2-HY^3_0 \dot Y^1_0=\pi_2=\pi_{02}+H\left(\dot Y^2_H-Y^3_0 \pi_{01}\right)\, .
\end{equation}
So, we find the PB
\begin{eqnarray}
\{Z^1(\sigma),{Z^2}'(\bar\sigma)\}&\cong&\{Y^1(\sigma),\pi_2(\bar\sigma)\}=H\left[Y^3_0(\sigma)-Y^3_0(\bar\sigma)\right]\delta_{2\pi}(\sigma-\bar\sigma)\, .
\end{eqnarray}
Note that $\delta_{2\pi}(\sigma-\bar\sigma)$ is $2\pi$ periodic $\delta$-function, $\delta_{2\pi}(\alpha)=\sum_{n\in Z}\delta(\alpha-2\pi n)$, so the periodic parts in bracket in front of $\delta$-function disappear and we obtain
\begin{equation}
\{Z^1(\sigma),Z'^2(\bar\sigma)\}=HN^3 (\sigma-\bar\sigma)\delta_{2\pi}(\sigma-\bar\sigma)\, .
\end{equation}
Here $N^3$ is winding number of $Y^3_0$ which has a general form
\begin{equation}
Y^3_0(\sigma)=N^3\sigma+Y^3_{periodic}(\sigma)\, .
\end{equation}
The expression $\alpha\delta_{2\pi}(\alpha)$ is zero for $\alpha=0$ but it is different from zero for $\alpha=2n\pi\,(n\in Z,n\neq 0)$.

The
integration over $\bar\sigma$, from $\bar\sigma_0$ to $\bar\sigma$, produces
\begin{equation}
\{Z^1(\sigma),Z^2(\bar\sigma)\}-\{Z^1(\sigma),Z^2(\bar\sigma_0)\}=-\frac{1}{2\pi}HN^3\left[F(\sigma-\bar\sigma)-F(\sigma-\bar\sigma_0)\right]\, ,
\end{equation}
where
\begin{equation}
2\pi \int_{\alpha_0}^\alpha d\eta \eta \delta_{2\pi}(\eta)=F(\alpha)-F(\alpha_0)\, ,
\end{equation}
and
\begin{equation}
F(\alpha)=\sum_{n\neq 0}\frac{1}{n^2}e^{-in\alpha}+i\alpha \sum_{n\neq 0} \frac{1}{n}e^{-in\alpha}+\frac{\alpha^2}{2}\, .
\end{equation}
The function $F(\alpha)$ is even $F(-\alpha)=F(\alpha)$
and
$F(0)=\frac{\pi^2}{3}$.

So, the result for PB itself
\begin{equation}\label{eq:PBz}
\{Z^1(\sigma),Z^2(\bar\sigma)\}=-\frac{1}{2\pi}HN^3\left[F(\sigma-\bar\sigma)+C\right]\, ,
\end{equation}
is in fact the equation (4.41) of Ref.\cite{ALLP} up to some integration constant $C$. The undetermined constant $C$ corresponds to the contribution of the zero modes of the undetermined commutators, because
one
started with $\sigma$-derivative of the coordinate $Z^{2}$. The choice of Ref.\cite{ALLP}
 in subsection 4.4.2  is
%corresponds to choice
 $C=0$ which produces the expression (4.41) of Ref.\cite{ALLP} and the noncommutativity at the same point $\sigma=\bar\sigma$
\begin{equation}
\{Z^1(\sigma),Z^2(\sigma)\}=-\frac{1}{2\pi}HN^3 F(0)=-\frac{\pi}{6}HN^3\, .
\end{equation}

As it was pointed out in Ref.\cite{ALLP}, "other reasonings could as well be pursued". Following the line of our paper one can require that coordinates are commutative at the same point ($\sigma=\bar\sigma$) which produces
\begin{equation}
C=-F(0)=-\frac{\pi^2}{3}\, .
\end{equation}
So, with this choice one has
\begin{equation}
\{Z^1(\sigma),Z^2(\bar\sigma)\}=HN^3\left[F(\sigma-\bar\sigma)-\frac{\pi^2}{3}\right]\, ,
\end{equation}
and obtains the non-commutativity for $\sigma=2\pi+\bar\sigma$
\begin{equation}
\{Z^1(\sigma+2\pi),Z^2(\sigma)\}=\pi HN^3\, .
\end{equation}

\subsection{Similarities and differences}

Although we analyzed the different cases, let us compare some general features of the results considered. In both approaches the commutators are infinitesimally small and they close on some winding numbers.
Note that in general, we can connect any geometric background with every nongeometric background from the chain of T-duality (\ref{eq:chain2}). Using the T-duality transformations we can calculate the noncommutativity of the coordinates of the nongeometric background in terms of the winding numbers of the geometrical background.

For arbitrary $\sigma$ and $\bar\sigma$, $\sigma$-dependence is different. In Ref.\cite{ALLP}, up to the integration constant $C$ it is equal to
 $$F(\sigma-\bar\sigma)+C\, ,$$ and
in the present article,
up to the integration constant $C_{1}$, it is
$$\left[x^\mu(\sigma)-x^\mu(\bar\sigma)\right]\theta(\sigma-\bar\sigma)+C_1\, .$$
The constants appear because in both approaches
we started with the sigma derivatives of the coordinates.
In the papers considered, the values of the constants are taken to be
$C=0$ and $C_{1}=0$.
For these choices, the noncommutativity appears for $\sigma=\bar\sigma$
in the Ref. \cite{ALLP} and for $\sigma=\bar\sigma+2\pi$
in the present article.
For the other choice
$C=-F(0)=-\frac{\pi^2}{3}$ and $C_{1}=0$,
in both cases the coordinate commute
at the same point $\sigma=\bar\sigma$
and have nontrivial PB for $\sigma=\bar\sigma+2\pi$.
%The fact that in Ref.\cite{lust} non-commutativity appears for $\sigma=\bar\sigma$ and in our case for $\sigma=\bar\sigma+2\pi$ follows from different choice of the integration constant in (\ref{eq:PBz}) and is not important.

The main difference
between two approaches is the origin of noncommutativity. The nontrivial boundary conditions given in Eq.(2.25) of Ref.\cite{ALLP} are the source of noncommutativity in that article. Because Ref.\cite{ALLP} does not consider $T_3$-dualization, $\beta^0_\mu$-functions (introduced in Eq.(\ref{eq:beta})) are zero and there is no noncommutativity of this kind. On the other hand, in the case considered in this paper, just these $\beta^0_\mu$ functions are the sources of the noncommutativity, even in the absence of the nontrivial boundary conditions of Ref.\cite{ALLP}. For complete noncommutativity relations one should take into account both kinds of noncommutativity.

%%%%%%%%%%%%%%%%%%%%%%%%%%%%%%%%%%%%%%%%%%%%%%%%%%
\section{Concluding remarks}
\setcounter{equation}{0}

In the present article we derived the closed string non-commutativity relations. We considered the theory describing the string moving in the weakly curved background. Its T-dual theory is obtained performing the T-dualization procedure along all the coordinates \cite{DS}. The T-dual transformation laws have the central role in our approach.
These laws connect the world-sheet derivatives of the coordinates and momenta in the original and the T-dual theory.
The zero orders are transformation laws of the constant background and they do not lead to the noncommutativity. The term $\beta^0_\mu$, which is infinitesimally small and bilinear in coordinates $x^\mu$, plays the key role in obtaining the noncommutativity relations.

In the original space we choose the standard Poisoon brackets. The T-dual coordinates $y_\mu$ has two terms: one linear in the original momenta and the other bilinear in the original coordinates. This explains the nontrivial PB $\{y_\mu,y_\nu\}$ (\ref{eq:yy}) which is linear in coordinate. Note that in the case of open string moving in the flat background coordinate is linear function in both effective momenta and coordinates. So, the corresponding PB is constant.

The T-dual momenta ${}^\star \pi^\mu$ are bilinear expressions in original coordinates. So, PB of the T-dual momenta vanishes (\ref{eq:pp}), but PB between T-dual coordinates and momenta (\ref{eq:ypi}) obtained additional term linear in coordinates.

In the doubled space there
exists the additional
coordinate $\tilde y_\mu$. It consists of
the term linear in original momenta, but with the coefficient linear
in original coordinate and the other terms bilinear in original
coordinates. So, it produces the nontrivial PB with all variables
$(y_\mu,\tilde y_\mu,{}^\star \pi^\mu)$, (\ref{eq:yty}), (\ref{eq:tyty}) and (\ref{eq:typi}).

The general structure of the non-commutativity relations is
\begin{equation}
\{Y_\mu(\sigma),Y_\nu(\bar\sigma)\}=\{F_{\mu\nu\rho}\left[x^\rho(\sigma)-x^\rho(\bar\sigma)\right]+\tilde F_{\mu\nu\rho}\left[\tilde x^\rho(\sigma)-\tilde x^\rho(\bar\sigma)\right]\}\theta(\sigma-\bar\sigma)\, ,
\end{equation}
where $Y_\mu=(y_\mu,\tilde y_\nu)$ and $F_{\mu\nu\rho}$ and $\tilde F_{\mu\nu\rho}$ are the constant and infinitesimally small fluxes. At the same points, for $\sigma=\bar\sigma$ all PB's are zero. In the important particular case for $\sigma=\bar\sigma+2\pi$ we get
\begin{equation}
\{Y_\mu(\sigma+2\pi),Y_\nu(\sigma)\}=2\pi \left[(F_{\mu\nu\rho}+2\tilde F_{\mu\nu\alpha}b^\alpha_\rho)N^\rho+\frac{1}{\kappa}\tilde F_{\mu\nu}{}^\rho p_\rho\right]\, ,
\end{equation}
where $N^\mu$ and $p_\mu$ are winding numbers and momenta of the original theory. We can
rewrite it in the form
\begin{equation}\label{eq:cro}
\{Y_\mu(\sigma+2\pi),Y_\nu(\sigma)\}=\oint_{C_\rho} F_{\mu\nu\rho}d x^\rho+\oint_{\tilde C_\rho} \tilde F_{\mu\nu\rho} d\tilde x^\rho\, ,
\end{equation}
where $C_\rho$ and $\tilde C_\rho$ are cycles around which the
closed string is wrapped. Note the "wrapping" of auxiliary
coordinate $\tilde x^\mu$ is in accordance with
(\ref{eq:wrapping}) and represents linear combination of momenta
$p_\mu$ and winding numbers $N^\mu$. This generalizes the
conjecture of Ref.\cite{ALLP3} between the closed string noncommutativity and fluxes.

In terms of Ref.\cite{ALLP} for the three dimensional torus $x^\mu\rightarrow X^{a},\,(a=1,2,3)$
our case corresponds to the non-commutativity of the nongeometric background
with $W^{a}$ coordinates and $R$-fluxes obtained after the
successive performation  of all three T-dualizations
along all three coordinates.
It relates $W^{a}$ with $X^{a}$ coordinates of torus with $H$-flux,
and so the PB closes on the winding number of the $X^{a}$-coordinates.
We hope that these results will contribute
to the better understanding of the most strange, uncommon R-flux configurations
where the noncommutativity appears as a consequence of the nontrivial $\beta^{0}_\mu$-functions.
Note that Ref.\cite{ALLP} uses $T_{2}$-duality
(performed along $Y^{2}$) and the relation $Z^{a}=Z^{a}(Y^{a})$
to obtain the non-commutativity of the nongeometric
background with $Q$-flux in terms of the winding of $Y^{a}$-coordinates.
There the noncommutativity originates from the nontrivial boundary conditions.
To obtain the general structure of the closed string noncommutativity
for arbitrary background of the chain (\ref{eq:chain2}) one should find its T-duality transformations
with all other backgrounds of the chain and
calculate both kind of the noncommutativity originating from
nontrivial boundary conditions as well as from nontrivial $\beta_\mu^{0}$ functions.

The term of the action with the constant part of the Kalb-Ramond field $b_{\mu\nu}$ is topological. So, it does not contribute to the equations of motion. In the open string case it contributes to the boundary conditions and it is a source of the open string noncommutativity. In the closed string case it is absent from boundary conditions as well. Classically, we can gauge it away and Kalb-Ramond field becomes infinitesimally small. But, if $b_{\mu\nu}=0$ one loses toplogical contributions. In order to investigate the global structure of the theory with holonomies of the world sheet gauge fields in quantum theory we should preserve such term.

Putting $b_{\mu\nu}=0$ the noncommutativity relations (\ref{eq:nc1}), (\ref{eq:nc2}) and (\ref{eq:nc3}) get the simpler form
\begin{eqnarray}
\{y_\mu(\sigma+2\pi), y_\nu(\sigma)\}&=&-\frac{2\pi}{\kappa}B_{\mu\nu\rho}N^\rho\, ,\nonumber\\ \{y_\mu(\sigma+2\pi),\tilde y_\nu(\sigma)\}&=&-\frac{1}{\kappa}G_{\mu\nu}-\frac{2\pi}{\kappa^2}B_{\mu\nu}{}^\rho p_\rho\, ,\\ \{\tilde y_\mu(\sigma+2\pi),\tilde y_\nu(\sigma)\}&=&-\frac{6\pi}{\kappa}B_{\mu\nu\rho}N^\rho\, .\nonumber
\end{eqnarray}

%%%%%%%%%%%%%%%%%%%%%%%%%%%%%%%%%%%%%%
\appendix

\section{The momentum in the T-dual theory}\label{sec:dm}
\cleq

Let us here calculate the T-dual momentum given in (\ref{eq:dualmom}).
The T-dual theory depends on two variables
$y_\mu$, $\tilde{y}_\mu$
which are connected by the relations $\dot{y}_\mu=\tilde{y}^\prime_\mu$, ${y}^\prime_\mu=\dot{\tilde{y}}_\mu$.
So, to obtain the momentum canonically conjugated to $y_\mu$,
we should vary the action with respect to both $\dot{y}_\mu$ and $\tilde{y}^\prime_\mu$.

First, let us calculate the contribution from the background fields argument.
With the help of the relation
\begin{equation}
\Theta^{\mu\nu}_{-}[x]=\Theta^{\mu\nu}_{0-}-2\kappa \Theta_{0-}^{\mu\rho}h_{\rho\sigma}[x]\Theta^{\sigma\nu}_{0-}\, ,
\end{equation}
we can rewrite the T-dual action (\ref{eq:dualna}) as
\begin{eqnarray}
{}^\star S[y]&=&
{}^\star S_0-
\kappa^3\int d^2\xi\, \partial_+y_\mu \Theta^{\mu\rho}_{0-}h_{\rho\sigma}
\big[\Delta  V[y]\big]
\Theta^{\sigma\nu}_{0-}\partial_- y_\nu\, ,
\nonumber\\
{}^\star S_0&=&\frac{\kappa^{2}}{2}
\int d^{2}\xi\
\partial_{+}y_\mu
\Theta_{0-}^{\mu\nu}
\partial_{-}y_\nu.
\end{eqnarray}
Using the expression
\begin{equation}
\partial_\pm V^{\mu}=-\kappa \Theta^{\mu\nu}_{0\pm} \partial_\pm y^{(0)}_\nu\, ,
\end{equation}
we obtain
\begin{equation}
{}^\star S[y]={}^\star S_0
+\kappa \int d^2\xi\, \partial_+ V^{\mu} h_{\mu\nu}\big[\Delta V\big]\partial_- V^{\nu}
={}^\star S_0+\kappa \int d^2\xi \Delta V^\mu h_{\mu\nu}\big[\partial_- V\big]
\partial_+ V^{\nu}.
\end{equation}
Because of the relation
\begin{equation}
h_{\mu\nu}\big[\partial_- V\big]
\partial_ +V^{\nu}
=\partial_0 \beta^0_\mu\big[V\big]+\partial_1 \beta^1_\mu\big[V\big]\, ,
\end{equation}
the action becomes
\begin{equation}\label{eq:da}
{}^\star S[y]={}^\star S_0+\kappa \int d^2\xi
\Big[\kappa \Delta y_\mu\theta^{\mu\nu}_{0}+\Delta\tilde y_\mu (g^{-1})^{\mu\nu}\Big]
\Big(\partial_0 \beta^0_\nu\big[V\big]+\partial_1 \beta^1_\nu\big[V\big]\Big)\, .
\end{equation}
%We write $y$ instead $y^{(0)}$ because it stands besides $\beta^1$ which is of the first order in $B_{\mu\nu\rho}$, and we do all calculations in the linear approximation in $B_{\mu\nu\rho}$.

So, the contribution to the T-dual momentum, coming from the T-dual background fields
argument is obtained from (\ref{eq:da}),
integrating over $\sigma$ by parts in
$\Delta \tilde y_\mu  (g^{-1})^{\mu\nu}\partial_1 \beta^1_\nu$.
Using
$\tilde y'_\mu=\dot y_\mu$
we obtain
\begin{equation}
\Delta\,^\star\pi^\mu=-\kappa  (g^{-1})^{\mu\nu}\beta^1_\nu\big[V\big].
\end{equation}
Therefore, the total T-dual momentum is
\begin{equation}
^\star\pi^\mu=
\kappa(G^{-1}_{E})^{\mu\nu}\big[\Delta  V[y]\big]
\dot{y}_\nu
-\kappa^{2}\theta^{\mu\nu}\big[\Delta  V[y]\big]
y^\prime_\nu
-\kappa  (g^{-1})^{\mu\nu}\beta^1_\nu\big[V[y]\big].
\end{equation}

\section{Fluxes}\label{sec:tdfs}
\cleq

The field strength of the original  Kalb-Ramond field, is given by (\ref{eq:fskr}).
The original metric $G_{\mu\nu}$ is constant, and therefore the corresponding Christoffel connection
is zero.
The effective metric $G^{E}_{\mu\nu}$ is linear  in coordinate and the corresponding Christoffel
connection
\begin{equation}\label{eq:Gammaeff}
\Gamma^{E}_{\mu,\nu\rho}=\frac{1}{2}\Big{(}
\partial_\nu G^{E}_{\mu\rho}+
\partial_\rho G^{E}_{\mu\nu}-
\partial_\mu G^{E}_{\nu\rho}\Big{)}=
-\frac{4}{3}\Big{(}
B_{\mu\sigma\nu}(G^{-1}b)^{\sigma}_{\ \rho}
+B_{\mu\sigma\rho}(G^{-1}b)^{\sigma}_{\ \nu}
\Big{)},
\end{equation}
is the infinitesimally small constant.
It will be used in the following forms
\begin{equation}\label{eq:Geff}
\Gamma^{E}_{\mu,\nu\rho}x^\mu=
4\Big{(}
h[x]G^{-1}b+bG^{-1}h[x]
\Big{)}_{\nu\rho},
\end{equation}
and
\begin{equation}
(\Gamma^{E}_{\mu,\nu\rho}-\Gamma^{E}_{\nu,\mu\rho})x^\rho=
8h_{\mu\nu}[bx]-4
\Big{(}
h[x]G^{-1}b-bG^{-1}h[x]
\Big{)}_{\mu\nu}.
\end{equation}

%%%%%%%%%%%%%%%%%%%%%%%%%%%%

We can express
the dual Kalb-Ramond field \cite{DS}
as
\begin{equation}
^\star B^{\mu\nu}[\Delta V]=\,^\star b^{\mu\nu}+Q^{\mu\nu}_{\ \ \rho}\Delta V^\rho,
\end{equation}
where $^\star b^{\mu\nu}=\frac{\kappa}{2}\theta^{\mu\nu}_{0}$ and
\begin{equation}\label{eq:qu}
Q^{\mu\nu}_{\ \ \rho}=
-\frac{1}{3}\Big{[}
(g^{-1})^{\mu\sigma}(g^{-1})^{\nu\tau}
-\kappa^{2}\theta^{\mu\sigma}_{0}\theta^{\nu\tau}_{0}
\Big{]}B_{\sigma\tau\rho}.
\end{equation}
This will be used as
\begin{eqnarray}
Q^{\mu\nu}_{\ \ \rho}x^\rho
&=&-(g^{-1})^{\mu\rho}\Big{[}h[x]
+4bG^{-1}h[x]G^{-1}b
\Big{]}_{\rho\sigma}(g^{-1})^{\sigma\nu}
\nonumber\\
&=&
-\Big{[}
g^{-1}h[x]g^{-1}
+\kappa^{2}\theta_{0}h[x]\theta_{0}
\Big{]}^{\mu\nu}.
\end{eqnarray}

\section{PB's between pure coordinates}\label{sec:purecoor}
\cleq

Starting with the PB of the $\sigma$ derivatives of the coordinates
\begin{equation}\label{eq:ypyp}
\{X^{\prime}_{\mu}(\sigma),Y^{\prime}_{\nu}(\bar\sigma)\}\cong
K^{\prime}_{\mu\nu}(\sigma)\delta(\sigma-\bar\sigma)
+L_{\mu\nu}(\sigma)\delta^\prime(\sigma-\bar\sigma),
\end{equation}
let us find the expression for the PB between coordinates
$\{X_{\mu}(\sigma),Y_{\nu}(\bar\sigma)\}$.
From (\ref{eq:ypyp}) it follows that $\Delta X_\mu(\sigma,\sigma_0)$ and $\Delta Y_\mu(\sigma,\sigma_0)$ defined by
\begin{eqnarray}
\Delta X_\mu(\sigma,\sigma_{0})&=&
\int_{\sigma_{0}}^\sigma d\eta\,X^{\prime}_{\mu}(\eta)=
X_\mu(\sigma)-X_\mu(\sigma_{0}),
\nonumber\\
\Delta Y_\mu(\sigma,\sigma_{0})&=&
\int_{\sigma_{0}}^\sigma d\eta\,Y^{\prime}_{\mu}(\eta)=
Y_\mu(\sigma)-Y_\mu(\sigma_{0}),
\end{eqnarray}
satisfy
\begin{equation}
\{\Delta X_{\mu}(\sigma,\sigma_0),\Delta Y_{\nu}(\bar\sigma,{\bar{\sigma}}_0)\}
\cong
\int_{\sigma_0}^\sigma d\eta
\int_{\bar\sigma_0}^{\bar{\sigma}}d\bar{\eta}
\Big{[}
 K'_{\mu\nu}(\eta)\delta(\eta-\bar{\eta})
+L_{\mu\nu}(\eta)\delta^\prime(\eta-\bar{\eta})
\Big{]}.
\end{equation}
Integrating over $\bar\eta$ and
using
\begin{equation}\label{eq:fint}
\int_{\sigma_{0}}^{\sigma}d\eta f(\eta)\delta(\eta-\bar{\sigma})=
f(\bar{\sigma})\big{[}
\theta(\sigma-\bar{\sigma})-\theta(\sigma_{0}-\bar{\sigma})
\big{]},
\end{equation}
 we obtain
\begin{eqnarray}
\{\Delta X_\mu(\sigma,\sigma_0),\Delta Y_\nu(\bar\sigma,\bar\sigma_0)\}
&\cong&\int_{\sigma_0}^\sigma d\eta
\Big{[}
 K'_{\mu\nu}(\eta)
\big{[}
\theta(\eta-\bar\sigma_0)-\theta(\eta-\bar\sigma)\big{]}
\nonumber\\
&&\quad\quad\,+L_{\mu\nu}(\eta)
\big{[}
\delta(\eta-\bar\sigma_0)-\delta(\eta-\bar\sigma)\big{]}
\Big{]},
\end{eqnarray}
where the function $\theta(\sigma)$ is defined as
\begin{equation}\label{eq:fdelt}
\theta(\sigma)\equiv \int_0^\sigma d\eta \delta(\eta)
=\frac{1}{2\pi}\big{(}
\sigma+2\sum_{n\geq 1}\frac{1}{n}\sin{n\sigma}
\big{)}
=\left\{\begin{array}{ll}
0 & \textrm{if $\sigma=0$}\\
1/2 & \textrm{if $0<\sigma<2\pi$}, \quad \sigma\in[0.2\pi].\\
1 & \textrm{if $\sigma=2\pi$} \end{array}\right .
\end{equation}

Integrating by parts over $\eta$ and using (\ref{eq:fint}) we get
\begin{eqnarray}\label{eq:dxdx}
&&\{\Delta X_\mu(\sigma,\sigma_0),\Delta Y_\nu(\bar\sigma,\bar\sigma_0)\}
\cong
\nonumber\\
&&K_{\mu\nu}(\sigma)
\big[\theta(\sigma-\bar{\sigma}_{0})
-\theta(\sigma-\bar{\sigma})\big]
-K_{\mu\nu}(\sigma_{0})
\big[\theta(\sigma_{0}-\bar{\sigma}_{0})-\theta(\sigma_{0}-\bar{\sigma})\big]
\nonumber\\
&&\,-K_{\mu\nu}(\bar{\sigma}_{0})
\big[
\theta(\sigma-\bar{\sigma}_{0})
-\theta(\sigma_{0}-\bar{\sigma}_{0})\big]
+K_{\mu\nu}(\bar\sigma)
\big[\theta(\sigma-\bar{\sigma})-\theta(\sigma_{0}-\bar{\sigma})\big]
\nonumber\\
&&\,
+L_{\mu\nu}(\bar\sigma_{0})
\big{[}\theta(\sigma-\bar\sigma_{0})
-\theta(\sigma_{0}-\bar\sigma_{0})\big{]}
-L_{\mu\nu}(\bar\sigma)\big{[}
\theta(\sigma-\bar\sigma)-\theta(\sigma_{0}-\bar\sigma)\big{]}.
\end{eqnarray}
Relation
\begin{equation}\label{eq:XY}
\{X_\mu(\tau,\sigma),Y_\nu(\tau,\bar\sigma)\}\cong-\left[K_{\mu\nu}(\sigma)-K_{\mu\nu}(\bar\sigma)+L_{\mu\nu}(\bar\sigma)\right]\theta(\sigma-\bar\sigma),
\end{equation}
solves (\ref{eq:dxdx}), up to additive constant.

For $X_\mu=Y_\mu$, the antisymmetry of the left hand side under the replacement
$\mu\leftrightarrow\nu$ and $\sigma\leftrightarrow\bar\sigma$, produces conditions
$L_{\mu\nu}=L_{\nu\mu}$
and $K_{\mu\nu}+K_{\nu\mu}=L_{\mu\nu}$.

%%%%%%%%%%%%%%%%%%%%%%%%%%%%%%%%%%%%

\end{document}